\pdfoutput=1
\documentclass[aps,prc,twocolumn,nofootinbib,floatfix]{revtex4-2}

\usepackage[T1]{fontenc}
\usepackage{amsmath}
\usepackage{amssymb}
\usepackage{bm}
\usepackage{graphicx}
\usepackage{xcolor}

\usepackage[colorlinks=false,pdfborder={0 0 1},hidelinks]{hyperref}

\usepackage{cleveref}

\begin{document}

\title{Relating Short Range Correlated Pair Abundances to Electron-Scattering Cross Section Ratios}

\author{Ilay Wischnevsky Shlush}
\author{Igor Korover}
\author{Eli Piasetzky}
\affiliation{School of Physics and Astronomy, Tel Aviv University, Tel Aviv, 6997845, Israel}


\begin{abstract}
We derive a coupled set of relations between inclusive and semi-exclusive nucleon-knockout electron-scattering cross-section ratios measured for any two nuclei in short-range-correlation (SRC) kinematics and the microscopic neutron-proton, (np-), proton-proton (pp-), and neutron-neutron (nn-) SRC pair populations in those nuclei. The derivation uses a factorized description of quasielastic scattering from SRC pairs and includes final-state interactions through nuclear transparency and single charge exchange. We show that the inclusive $(e,e’)$, proton-knockout $(e,e'p)$, and neutron-knockout $(e,e'n)$ ratios are not independent: for any two nuclei where one is isospin symmetric ($N=Z$), two of these three observables determine the third within the model. For $^\mathrm{48}\mathrm{Ca}$ and $^\mathrm{40}\mathrm{Ca}$, for example, the measured inclusive $(e,e')$ and proton-tagged $(e,e'p)$ ratios predict $^\mathrm{48}\mathrm{Ca}/^\mathrm{40}\mathrm{Ca}(e,e'n)=1.35\pm0.07\,(\mathrm{exp})\pm0.15\,(\mathrm{model})$. A measurement of the $(e,e'n)$ ratio therefore provides a direct consistency test of the factorized pair-counting description. It also constrains the single-charge-exchange probabilities. Combining all three ratios with the two double-coincidence ratios determines the relative np-, pp-, and nn- SRC pair populations of any two nuclei. If one nucleus is isospin symmetric, any two of the three ratios suffice for the same result.
\end{abstract}

\maketitle

\section{Introduction}
\label{sec:intro}

Short-range correlations (SRCs) are transient configurations of two nucleons
separated by approximately $1$~fm or less, where the short-range repulsive
core and tensor components of the nucleon--nucleon (NN) interaction
dominate~\cite{HenRMP2017, Piasetzky2023}. These correlations generate the high-momentum
tail ($k > k_F \approx 250$~MeV/$c$) of the nuclear momentum distribution and
involve approximately 20\% of nucleons in medium-mass
nuclei~\cite{Subedi2008,Egiyan2003,Fomin2012}.

Electron-scattering measurements have established that SRC pairs are dominated by neutron-proton (np) configurations, whereas proton-proton (pp) and neutron-neutron (nn) pairs make substantially smaller contributions ~\cite{Subedi2008,Piasetzky2006,Duer2019PRL,Korover2014}. Consequently, in neutron-rich nuclei, protons carry a disproportionate fraction of the high-momentum strength ~\cite{Hen2014,Duer2018}. These observations have implications for the interpretation of neutrino-nucleus reactions and for the density dependence of the nuclear symmetry energy ~\cite{Schmookler2019}. Because neutron-skin observables probe the same underlying isovector nuclear dynamics, SRC effects may also be relevant to their interpretation in neutron-rich nuclei~\cite{Adhikari2022,Lucas}.

The dependence of SRC-pair formation on nuclear shell structure remains less well constrained. In the independent-particle shell model, $^{40}$Ca and  $^{48}$Ca are both doubly magic nuclei, with the eight additional neutrons in  $^{48}$Ca-corresponding to a $40\%$ increase in neutron number—primarily occupying the $1f_{7/2}$ orbital. An inclusive  $^{48}$Ca/$^{40}$Ca measurement found a $16.5\%$ increase in the per-nucleon cross-section ratio in SRC kinematics~\cite{Nguyen2020}. However, an inclusive observable alone cannot separate the contributions of (np), (pp), and (nn) SRC pairs~\cite{Weiss2021}. The CaFe experiment subsequently measured semi-exclusive proton-knockout cross sections from $^{40}$Ca, $^{48}$Ca, and $^{54}$Fe and observed a strong variation with the filling of the $1f_{7/2}$ neutron shell~\cite{CaFe1,CaFe2}. A framework combining inclusive and semi-exclusive reaction channels is therefore required to relate the measured cross-section ratios to the underlying SRC-pair populations.

We develop such a framework here. Within a factorized description of quasielastic scattering from SRC pairs, we derive a coupled set of relations connecting inclusive and semi-exclusive proton- and neutron-knockout cross-section ratios to the microscopic (np), (pp), and (nn) pair populations. Final-state interactions are incorporated through nuclear transparency and single charge exchange (SCX). A central result is that the inclusive, $(e,e'p)$, and $(e,e'n)$  ratios are not independent: for any two nuclei, two of these observables determine the third within the model. Testing this relation experimentally therefore provides a direct test of the approximations underlying the framework. We also identify the additional measurements needed to constrain SCX and determine the relative isospin composition of the SRC-pair population.
Recently, electron-induced $(e,e')$, $(e,e'p)$, $(e,e'n)$,  $(e,e'pp)$, and $(e,e'pn)$ cross-section ratios in SRC kinematics were measured for $^{48}$Ca relative to $^{40}$Ca~\cite{CaFe1}. Below, we use the published $(e,e')$ and $(e,e'p)$ results to predict the corresponding $(e,e'n)$ ratio. Once the remaining measured ratios become available, their combined analysis within the factorized SRC framework developed here will make it possible to constrain the relative (np), (pp), and (nn) pair populations in the two calcium isotopes and to investigate how SRC-pair formation changes with neutron-shell occupancy.

\section{Factorized SRC Cross Section Model}
\label{sec:formalism}

\subsection{Definitions and Assumptions}
\label{sec:assumptions}

Let $N_{np}^{A}$, $N_{pp}^{A}$, and $N_{nn}^{A}$ denote the numbers of
SRC-correlated $np$, $pp$, and $nn$ pairs in nucleus $A$.
The quantities $\sigma_{p}$ and $\sigma_{n}$ are the effective elementary electron-proton and electron-neutron cross sections in the selected kinematics. Nuclear transparencies are $T_p^A,T_n^A$ and $T_{2N}^A$ for detected proton, neutron, and two-nucleon final states. The probability $P_A$ describes a single charge exchange in nucleus A. Superscript B denotes the comparison nucleus.

The derivation assumes that the short-distance structure of a pair is universal across nuclei~\cite{HenRMP2017,Weiss2015,CruzTorres2021} and that the virtual photon couples to one nucleon in the pair. An np pair contributes $\sigma_p$ to proton knockout and $\sigma_n$ to neutron knockout. The two identical nucleons in pp or nn pairs produce factors of two. Transparency removes events from a tagged yield, while SCX transfers events between proton- and neutron-tagged samples~\cite{Korover2021,Duer2019PLB,Frankfurt2008}. Terms requiring two charge exchanges are of order $P^2$ and are neglected. 

The equations contain only two-nucleon SRC contributions. Mean-field events shifted into the sample by final-state interactions and three-nucleon SRC configurations are assumed negligible in the kinematic region $x_B \geq 1.2$, $Q^2 \geq 1.5   \rm{(GeV/c^2)}$ , and $375\leq p_{miss} \leq 700 \rm{MeV/c}$~\cite{CaFe2,Nguyen2020,Fomin2012}. The comparison of measurements taken in different acceptances further assumes that they sample the same underlying two-nucleon SRC configurations.

\subsection{Coupled Cross Section Ratios}
The target $B$-to-target $A$ cross section ratios read as follows, where $\Sigma_{p}^{A}\equiv N_{np}^{A}+2N_{pp}^{A}$ and $\Sigma_{n}^{A}\equiv N_{np}^{A}+2N_{nn}^{A}$.

\begin{widetext}
\begin{align}
R_{B/A}\left(e,e'\right)
&=\frac{B\left(e,e'\right)}{A\left(e,e'\right)}
=\frac{\Sigma_p^B\sigma_p+\Sigma_n^B\sigma_n}{\Sigma_p^A\sigma_p+\Sigma_n^A\sigma_n},
\label{eq:Ree}
\\
R_{B/A}\left(e,e'p\right)
&=\frac{B\left(e,e'p\right)}{A\left(e,e'p\right)}
=\frac{\Sigma_{p}^{B}\sigma_{p}T_{p}^{B}
+\Sigma_{n}^{B}\sigma_{n}T_{p}^{B}P_B}
{\Sigma_{p}^{A}\sigma_{p}T_{p}^{A}
+\Sigma_{n}^{A}\sigma_{n}T_{p}^{A}P_A},
\label{eq:Reep}
\\
R_{B/A}\left(e,e'n\right)
&=\frac{B\left(e,e'n\right)}{A\left(e,e'n\right)}
=\frac{\Sigma_{n}^{B}\sigma_{n}T_{n}^{B}
+\Sigma_{p}^{B}\sigma_{p}T_{n}^{B}P_B}
{\Sigma_{n}^{A}\sigma_{n}T_{n}^{A}
+\Sigma_{p}^{A}\sigma_{p}T_{n}^{A}P_A},
\label{eq:Reen}
\\
R_{B/A}\left(e,e'np\right)
&=\frac{B\left(e,e'np\right)}{A\left(e,e'np\right)}
=\frac{\left(N_{np}^{B}+2N_{nn}^{B}P_B\right)\sigma_{n}T_{2N}^{B}
+2N_{pp}^{B}P_B\sigma_{p}T_{2N}^{B}}
{\left(N_{np}^{A}+2N_{nn}^{A}P_A\right)\sigma_{n}T_{2N}^{A}
+2N_{pp}^{A}P_A\sigma_{p}T_{2N}^{A}},
\label{eq:Reenp}
\\
R_{B/A}\left(e,e'pp\right)
&=\frac{B\left(e,e'pp\right)}{A\left(e,e'pp\right)}
=\frac{2N_{pp}^{B}\sigma_{p}T_{2N}^{B}+N_{np}^{B}\sigma_{n}P_BT_{2N}^{B}}
{2N_{pp}^{A}\sigma_{p}T_{2N}^{A}+N_{np}^{A}\sigma_{n}P_AT_{2N}^{A}}.
\label{eq:Reepp}
\end{align}

The same-target double-to-single ratios provide composition information that target-to-target ratios alone do not supply:

\begin{align}
R_{p}^{A}
&=\frac{A\left(e,e'pp\right)}{A\left(e,e'p\right)}
=\frac{2N_{pp}^{A}\sigma_{p}T_{2N}^{A}+N_{np}^{A}\sigma_{n}P_AT_{2N}^{A}}
{\Sigma_p^A\sigma_{p}T_{p}^{A}
+\Sigma_n^A\sigma_{n}T_{p}^{A}P_A},
\label{eq:RpA}
\\
R_{n}^{A}
&=\frac{A\left(e,e'np\right)}{A\left(e,e'n\right)}
=\frac{\left(N_{np}^{A}+2N_{nn}^{A}P_A\right)\sigma_{n}T_{2N}^{A}
+2N_{pp}^{A}P_A\sigma_{p}T_{2N}^{A}}
{\Sigma_n^A\sigma_{n}T_{n}^{A}
+\Sigma_p^A\sigma_{p}T_{n}^{A}P_A}.
\label{eq:RnA}
\end{align}

\end{widetext}

Equations~\eqref{eq:Ree}--\eqref{eq:RnA}  are written for acceptance and detection corrected yields. When the cross sections are already corrected for nuclear transparency, the corresponding transparency factors must be removed consistently from the equations. The inclusive ratio is unaffected. In the analysis below, all tagged inputs and predictions use the transparency-corrected convention, leaving SCX as the only explicit final-state-interaction parameter.

\section{CONTENT OF THE COUPLED SYSTEM}

\subsection{Dependence of inclusive and single-knockout ratios}
\label{sec:dependence}

Equations \eqref{eq:Ree}--\eqref{eq:Reen} depend on the same two pair-weighted combinations, $\Sigma_p$ and $\Sigma_n$. Consequently, in the case of symmetric A, the inclusive and semi-exclusive proton- and neutron-knockout ratios contain only two independent constraints. For any pair of nuclei $(A,B)$ such that A is symmetric, measuring any two ratios determines the third, once $\rho\equiv\sigma_p/\sigma_n$, the SCX probabilities, and the transparency convention are specified. A third measurement therefore tests the assumptions of factorization, two-nucleon SRC dominance, and the treatment of final-state interactions; it does not add an independent pair-counting constraint.
For transparency-corrected ratios, elimination of the pair populations gives

\begin{widetext}
\begin{equation}
R(e,e'n)=\frac{\left(1+P_B\right)\left(1+\frac{\sigma_{p}}{\sigma_{n}}\right)R(e,e')-
\left(\frac{\sigma_{p}}{\sigma_{n}}+P_A\right)R(e,e'p)}
{1+P_A\frac{\sigma_{p}}{\sigma_{n}}},
\label{eq:consistency}
\end{equation}
\end{widetext}

Equation \eqref{eq:consistency} is the consistency relation used below. Cyclic rearrangement permits the prediction of either the inclusive or proton-knockoutd ratio if the other two are measured.

\subsection{Measurement hierarchy}
The coupled system separates tests of the reaction model from the extraction of pair populations. Any two observables among Eqs. \eqref{eq:Ree}--\eqref{eq:Reen} predict the third; measuring all three therefore provides a closure test and can constrain SCX when the elementary cross-section ratio and transparency convention are controlled.

To determine the target-to-target ratios of the np-, pp-, and nn- SRC populations, two independent observables from Eqs. \eqref{eq:Ree}--\eqref{eq:Reen} must be combined with both double-coincidence ratios in Eqs. \eqref{eq:Reenp} and \eqref{eq:Reepp}. A same-target double-to-single ratio, Eq. \eqref{eq:RpA} or Eq. \eqref{eq:RnA}, then removes the remaining composition degeneracy.

For an isospin-symmetric reference nucleus, the additional condition $N_{nn}^A = N_{pp}^A$ fixes the full relative pair content of both nuclei. These constraints determine ratios or relative compositions; one external normalization is still required to obtain absolute pair numbers.

\section{APPLICATION TO CALCIUM ISOTOPES}

\subsection{Experimental inputs}
We apply the framework to B = $^{48}$Ca and A = $^{40}$Ca. The inclusive ratio was measured by experiment E08-014 in Hall A, and the proton-tagged ratio by the CaFe experiment E12-16-004 in Hall C ~\cite{Nguyen2020,CaFe1,CaFe2}. The tagged ratio is corrected for nuclear transparency. Because the measurements were performed at different kinematics, their combination assumes SRC universality over both acceptances.

\begin{table}[htb]
\caption{Measured model inputs used for the $^{48}\mathrm{Ca}/^{40}\mathrm{Ca}$ calculation.
}
\label{tab:inputs}
\begin{ruledtabular}
\begin{tabular}{lll}
Quantity & Value & Source\\
\hline
$R_{48/40}(e,e')$ & $1.165 \pm 0.014$ & ~\cite{Nguyen2020}\\
$R_{48/40}(e,e'p)$ & $1.10 \pm 0.02$ & ~\cite{CaFe1,CaFe2}\\[0.4em]
$P_{40}$ & $0.04 \pm 0.03$ & ~\cite{Denniston2026,Friedes1965}\\
$P_{48}$ & $0.05 \pm 0.03$ & ~\cite{Denniston2026,Friedes1965}\\
$\rho = \sigma_{p}/\sigma_{n}$ & $2.5 \pm 0.5$ & ~\cite{deForest1983,Kelly2004}\\[0.4em]
\end{tabular}
\end{ruledtabular}
\end{table}

\subsection{Prediction for neutron knockout}
Substitution of the measured inclusive and proton-knockout  ratios into Eq. \eqref{eq:consistency} gives

\begin{equation}
R(e,e'n)_{\mathrm{pred}}=1.35\pm0.07\,(\mathrm{exp})\pm0.15\,(\mathrm{model}),
\label{eq:consistency_check}
\end{equation}

The experimental uncertainty propagates the errors on $R_{48/40}(e,e')$ and $R_{48/40}(e,e'p)$. The model uncertainty propagates the uncertainties in $P_{40},P_{48}$ and $\rho$. No neutron-knockout datum enters Eq. \eqref{eq:consistency_check}. A measurement of $R_{48/40}(e,e'n)$ in matched SRC kinematics is therefore a direct test of the common pair populations assumed by Eqs. \eqref{eq:Ree}--\eqref{eq:Reen}. Agreement would support the factorized description and the adopted treatment of SCX; disagreement would identify a discrepancy, but cannot distinguish between broken factorization, missing reaction mechanisms, mismatched kinematic sampling, or an incomplete final-state-interaction model.

\subsection{Same target ratios and SCX}
For an isospin-symmetric nucleus, charge symmetry gives $N_{nn}^A = N_{pp}^A$. If additionally the proton and neutron transparencies are equal, the elementary-cross-section-normalized same-target ratio becomes independent of the pair numbers:

\begin{equation}
R_{p/n}^{A}
=\frac{A\left(e,e'p\right)/\sigma_{p}}{A\left(e,e'n\right)/\sigma_{n}}
=\frac{\rho+P_A}{\rho{\left(1+\rho P_A\right)}}
\label{eq:Rpn}
\end{equation}

For $^{40}$Ca, $\rho=2.5\pm0.5$ and $P_{40}=0.04\pm0.03$ give $R_{p/n}^{40}=0.92\pm0.05$. The target-to-target relation $R_{p/n}^{B}=R_{p/n}^{A}\,
\frac{R_{B/A}\left(e,e'p\right)}{R_{B/A}\left(e,e'n\right)}$ together with equation \eqref{eq:consistency_check} gives $R_{p/n}^{48}=0.75\pm0.05\rm{(exp)}\pm0.07\rm{(model)}$.

The logic can also be reversed. A measurement of $R_{p/n}^{40}$ determines $P_{40}$ without reference to any pair population:

\begin{equation}
P^{A}=\frac{\rho\left(1-R_{p/n}^{A}\right)}{\rho^{2}R_{p/n}^{A}-1},
\label{eq:PAextract}
\end{equation}

Since
$\left| \partial R_{p/n}^{A}/\partial P^{A} \right| \approx1.7$ at these kinematics, a
measurement of $R_{p/n}^{A}$ to $\pm\Delta R^A_{p/n}$ determines $P^{A}$ to
$\Delta P^A\approx \pm 0.6\Delta R^A_{p/n}\pm0.013$, where the additional $\pm0.013$ comes from propagation of the elementary cross section ratio uncertainty
$\rho$.

Assuming $\Delta R_{p/n}^A=0.02$ we obtain $\frac{\Delta P^A}{P^A}\approx44\% $ -- better than the $75\%$ of the present calculation. 

This estimate assumes the relatively large uncertainty $\rho=2.5\pm0.5$. A more precise determination of $\rho$ would reduce the corresponding model contribution and further improve the precision with which $P_A$ can be extracted.

Once $P_{40}$ is known, a measurement of $R_{p/n}^{48}$ fixes $P_{48}$ through the coupled target ratios. Measuring all three ratios in Eqs. \eqref{eq:Ree}--\eqref{eq:Reen} supplies an over constrained determination of SCX and simultaneously checks whether a common pair-counting solution exists.

\section{EXTRACTION OF ISOSPIN PAIR POPULATIONS}

Given sufficient data, equations \eqref{eq:Ree}--\eqref{eq:RnA} can be used to extract the abundances of SRC pairs of different isospin content in nuclei A, B. 

The inclusive and semi-exclusive observables constrain $\Sigma_p$ and $\Sigma_n$, not $N_{np}$, $N_{pp}$, and $N_{nn}$ separately. The two target-to-target double-coincidence ratios in Eqs. \eqref{eq:Reenp} and \eqref{eq:Reepp} supply sensitivity to pp- and np-tagged final states needed to resolve them individually. 

The five target-to-target ratios, Eqs. \eqref{eq:Ree}--\eqref{eq:Reepp}, fix both the internal $np:pp:nn$ composition of each nucleus and the relative abundances between nuclei A and B, but not the \textit{total} abundances in both. Additional ratio measurements would not solve this, as the ratios are insensitive to the total SRC abundance. Instead, they would serve as consistency checks for the model.

As discussed in section \ref{sec:dependence}, if the reference nucleus is isospin symmetric, $N_{nn}^A$ = $N_{pp}^A$ provides the complementary charge-symmetry constraint, resulting in the linear dependence of Eqs. \eqref{eq:Ree}--\eqref{eq:Reen}. In this case, any two independent observables among Eqs. \eqref{eq:Ree}--\eqref{eq:Reen}, combined with both Eqs. \eqref{eq:Reenp} and \eqref{eq:Reepp}, determine the complete relative pair content of both nuclei again.

The word relative is essential: ratios of cross sections are invariant under a common rescaling of all pair numbers. Determining absolute values of $N_{np}$, $N_{pp}$, and $N_{nn}$ therefore requires one absolute pair normalization or an independently normalized cross section.
In practice, the extraction should be performed as a joint fit to all available ratios rather than by sequential algebraic inversion. A joint likelihood naturally handles correlated experimental uncertainties, uncertainties in $\rho$ and the SCX probabilities, and the redundancy of Eqs. \eqref{eq:Ree}-\eqref{eq:Reen}. The fit residual for the third inclusive or semi-exclusive ratio then becomes a quantitative goodness-of-fit test of the reaction model.

\section{MODEL SCOPE AND EXPERIMENTAL TESTS}

The predictive relation is useful because its failure modes are identifiable. The most incisive test is a neutron-knockout measurement performed in kinematics matched to the inclusive and proton-knockout data. The analysis also requires a consistent definition of transparency-corrected yields and identical treatment of acceptance, radiative corrections, and elementary electron-nucleon cross sections. Residual mean-field strength, three-nucleon correlations, multistep rescattering, or charge exchange beyond first order would break the closure relation.
The present uncertainty on the calcium prediction is dominated by model inputs, particularly SCX and the elementary cross-section ratio. Direct same-target proton-to-neutron measurements would reduce this dependence by determining SCX from the data. Measurements of the double-coincidence ratios would then convert the framework from a consistency test into an extraction of the isospin-resolved SRC pair populations.

\section{CONCLUSION}

We derived a factorized connection between SRC pair populations and inclusive, semi-exclusive single-nucleon-knockout, and two-nucleon-knockout electron-scattering ratios. The inclusive, proton-knockout, and neutron-knockout ratios form an over constrained set: any two predict the third. For $^{48}$Ca relative to $^{40}$Ca, existing inclusive and proton-knockout measurements predict $R_{48/40}(e,e'n) = 1.35 \pm 0.07 \rm{(exp)} \pm 0.15 \rm{(model)}$. Measuring this ratio would directly test whether the same pair populations describe all three channels.
A complete measurement program has a clear hierarchy. All three inclusive and semi-exclusive ratios constrain SCX. Two independent ratios from this set, together with the np- and pp-tagged double-coincidence ratios, determine the pair-population ratios between nuclei. Adding one same-target double-to-single ratio fixes the relative isospin composition; an isospin-symmetric reference nucleus supplies the charge-symmetry constraint. An external normalization remains necessary to convert these relative populations into absolute pair numbers.

\section*{Acknowledgments}
This research was supported by the Israeli Science Foundation (ISF) under Grants No. 917/20 (E.P.), No. 371/25 (E.P.), and 830/24 (I.K.), the ISF NSFC joint research program Grant No. 3107/23 (E.P.), the BSF 2024036 grant (E.P.), and the PAZY Foundation Grants No. 520/23 (E.P.) and No. 737/25 (I.K.)

\end{document}